\documentclass[aps,prl,showpacs,twocolumn,amsmath,10pt,superscriptaddress,floatfix,nofootinbib,a4paper]{revtex4-1}
\usepackage{epsfig,amssymb,amsfonts,bm,color}

\newcommand{\be}{\begin{equation}}
\newcommand{\ee}{\end{equation}}
\newcommand{\bea}{\begin{eqnarray}}
\newcommand{\eea}{\end{eqnarray}}
\newcommand{\beas}{\begin{eqnarray*}}
\newcommand{\eeas}{\end{eqnarray*}}
\newcommand{\ds}{\displaystyle}
\newcommand{\vep}{{\bm p}}
\newcommand{\vek}{{\bm k}}
\newcommand{\veq}{{\bm q}}
\newcommand{\tv}{t^v}
\newcommand{\tw}{t^w}

\newcommand{\dq}{d^3q}
\newcommand{\Br}{\mbox{Br}}
\newcommand{\g}{S}

\begin{document}

\title{A practical parametrization for line shapes of near-threshold states}

\author{C. Hanhart}
\affiliation{Forschungszentrum J\"ulich, Institute for Advanced Simulation, Institut f\"ur Kernphysik (Theorie) and 
J\"ulich Center for Hadron Physics, D-52425 J\"ulich, Germany}

\author{Yu. S. Kalashnikova}
\affiliation{Institute for Theoretical and Experimental Physics, 117218,
B.Cheremushkinskaya 25, Moscow, Russia}

\author{P. Matuschek}
\affiliation{Forschungszentrum J\"ulich, Institute for Advanced Simulation, Institut f\"ur Kernphysik (Theorie) and 
J\"ulich Center for Hadron Physics, D-52425 J\"ulich, Germany}

\author{R. V. Mizuk}
\affiliation{Institute for Theoretical and Experimental Physics, 117218,
B.Cheremushkinskaya 25, Moscow, Russia}
\affiliation{National Research Nuclear University MEPhI, 115409, Moscow, Russia}
\affiliation{Moscow Institute of Physics and Technology, 141700, Dolgoprudny, Moscow Region, Russia}

\author{A. V. Nefediev}
\affiliation{Institute for Theoretical and Experimental Physics, 117218,
B.Cheremushkinskaya 25, Moscow, Russia}
\affiliation{National Research Nuclear University MEPhI, 115409, Moscow, Russia}
\affiliation{Moscow Institute of Physics and Technology, 141700, Dolgoprudny, Moscow Region, Russia}

\author{Q. Wang}
\affiliation{Forschungszentrum J\"ulich, Institute for Advanced Simulation, Institut f\"ur Kernphysik (Theorie) and 
J\"ulich Center for Hadron Physics, D-52425 J\"ulich, Germany}

\begin{abstract}
Numerous quarkonium(like) states lying near $S$-wave thresholds are observed experimentally.We propose a
self-consistent approach to these near-threshold states compatible with unitarity and analyticity.The underlying
coupled-channel system includes a bare pole and an arbitrary number of elastic and inelastic channels treated fully
nonperturbatively.The resulting analytical parametrization is ideally suited for a combined analysis of the data
available in various channels that is exemplified by an excellent overall description of the data for the charged
$Z_b(10610)$ and $Z_b(10650)$ states.
\end{abstract}

\pacs{14.40.Rt, 14.40.Pq, 11.55.Bq, 12.38.Lg}

\maketitle

\section{Introduction}

At present there is no doubt that QCD is the true theory of strong interactions, at least at the energy scale
presently accessible for experimental investigations. One of the remarkable features of QCD is the prediction of the
existence of multiconstituent states, with a structure more complex than just quark-antiquark or three-quark
configurations, which are conventionally referred to as ``exotic'' hadrons. Experimental searches and theoretical
studies of such exotic states constitute an important tool in investigations of nature. 
Since the discovery of the charmonium(like)\footnote{We refer to hadrons as to ``charmonium(like)'' or
``bottomonium(like)'' if they contain $c\bar{c}$ or $b\bar{b}$ quark-antiquark pair, respectively, however may have
extra constituents like light-quark pairs.} state $X(3872)$ in 2003 \cite{Choi:2003ue}, numerous experiments continue to
deliver intriguing data on other charmonium(like) and bottomonium(like) states lying above the
respective open-flavor thresholds. Although for most of these states it is not possible at present
to make definite conclusions concerning their nature, some of these states share an important {feature}, namely
they reside in the vicinity of strong $S$-wave thresholds and they are seen in both open-flavor
(elastic) and hidden-flavor (inelastic) final states. Paradigmatic examples of such states are the $X(3872)$ near the
$D^0\bar{D} ^{*0}$ threshold, the $Z_b(10610)$ and $Z_b(10650)$ near the $B^{(*)}\bar{B} ^*$ thresholds, and
the $Z_c(3900)$
and $Z_c(4020)/ Z_c(4025)$ near the $D^{(*)}\bar{D}^*$ thresholds. Since vast and detailed information is becoming
available from existing experiments, and even more precise data
are expected from future high-statistics and high-precision experiments
\cite{Abe:2010gxa,Drutskoy:2012gt,Asner:2008nq,Lutz:2009ff} for states that are already known (see
\cite{Brambilla:2010cs,Brambilla:2014jmp} for recent reviews) as well as
for ones that are new and as yet unobserved, adequate theoretical tools for the data analysis are urgently called
for. The aim of
this Letter is to propose such a tool that is especially useful in describing near-threshold phenomena. 

The traditional way to perform an analysis of the experimental data is by using of individual Breit-Wigner distributions
for each peak combined with suitable background functions. However, such an approach provides
only limited information on the states studied, since the Breit-Wigner parameters are reaction-dependent and the naive
algebraic sum of the Breit-Wigner distributions violates unitary. In addition, by analyzing each reaction channel
individually, 
one does not exploit the full information content provided by the measurements. The
approach proposed in this Letter provides an important link between various models and
first-principles calculations in QCD (for example, lattice simulations) from one side to the experimental data on the
other side. To this end, we build a
model-independent parametrization for near-threshold phenomena consistent with requirements from unitarity and
analyticity. The formulas derived allow one to perform a simultaneous analysis of the entire bulk of data for all
decay channels for given near-threshold states(s). The resulting parametrization includes in a fully
nonperturbative way a bare pole and an arbitrary number of elastic and inelastic channels. With the
help of not-very-restrictive and phenomenologically justifiable assumptions the formulas can 
be solved analytically, which makes them as ideal for data analysis. 
The parameters of the final expressions are renormalized quantities with a direct physical meaning.
The suggested parametrization is, therefore,
expected to have a broad impact on the analysis of experimental data and to provide important
insights into the phenomenology of the strong
interactions. 

\section{Parametrization for the line shapes}

We consider a coupled-channel approach based on the Lippmann-Schwinger equation (LSE) for the $t$ matrix $t$,
\be
t=\hat{V}-\hat{V}\g t,
\ee
where  $\g$ denotes the free propagator in the corresponding channel. 
The potential 
\be
\raisebox{-3mm}{$\hat{V}=$}
\begin{tabular}{cc}
$\scriptstyle\mbox{\scriptsize Pole}\hspace*{7mm}\beta=\overline{1,N_e}\hspace*{7mm}i=\overline{1,N_{in}}$&\\[2mm]
$
\begin{pmatrix}
0&f_\beta(\vep')&f_i(\vek)\\[2mm]
f_\alpha(\vep)~&~v_{\alpha\beta}(\vep,\vep')~&~v_{\alpha i}(\vep,\vek)\\[2mm]
f_j(\vek')&v_{j\beta}(\vek',\vep')&v_{ji}(\vek',\vek)
\end{pmatrix}
$&\hspace*{-2mm}
\begin{tabular}{c}
\scriptsize Pole\\[2mm]
\scriptsize$\alpha=\overline{1,N_e}$\\[2mm]
\scriptsize$j=\overline{1,N_{in}}$.
\end{tabular}
\end{tabular}
\label{Vpot}
\ee
contains all possible types of interaction between the bare pole (labeled as ``0''---for
example, its position is $M_0$), the set of $N_e$ elastic open-flavor channels $(Q\bar{q})(q\bar{Q})$ (here $Q$ and $q$
denote a heavy and a light quark, respectively) labeled by Greek letters, and a set of
$N_{in}$ inelastic hidden-flavor channels $(Q\bar{Q})(q\bar{q})$ referred to by 
Latin letters. 

In order to proceed with the analytic solution, we make a few simplifying assumptions. In general
there are good reasons to neglect the direct interactions in the inelastic channels. For example, for the $X(3872)$
transitions between the $\rho J/\psi$ and $\omega J/\psi$ channels are forbidden by the isospin
conservation. In addition, since there are no light quarks inside the $J/\psi$ state, the direct
potential for $\rho (\omega) J/\psi\to \rho (\omega) J/\psi$ is also expected to be weak.
Analogously, since there are no light quarks in the heavy quarkonia $\Upsilon(nS)$ and $h_b(mP)$,
their interaction with pions is expected to be weak, with obvious relevance for the $Z_b^{(\prime)}$ 
states. Indeed, effective field theory estimates \cite{Liu:2012dv} and lattice calculations \cite{Detmold:2012pi} give
very small values for the scattering lengths of the pion scattered off the $c\bar{c}$ and $b\bar{b}$ quarkonia.
We therefore set $v_{ji}(\vek',\vek)=0$.
Next, we assume a separable form of the elastic transition vertex\footnote{A microscopic model for this interaction can
be found, for example, in \cite{Danilkin:2011sh,Danilkin:2009hr}.}, $v_{\alpha
i}(\vep,\vek)=\chi_\alpha(\vep)\varphi_{i\alpha}(\vek)$, where the additional assumption was made that
$\chi_{i\alpha}$ is independent of $i$. Indeed, the transition of the open-flavor channels to the hidden-flavor
channels demands the exchange of a heavy meson and,
therefore, it is necessarily of short range for all inelastic channels. Without loss of generality we set
$\chi_\alpha(\vep=0)=1$.
In addition, in a relatively narrow region near the elastic threshold(s) it is sufficient to parametrize the transition
form factors as
$$
f_\alpha(\vep)=f_\alpha,\chi_\alpha(\vep)=1,\varphi_{i\alpha}(\vek)=g_{i\alpha}|\vek|^{l_i},
f_i(\vek)=\lambda_i |\vek|^{l_i},
$$
where $f_\alpha$, $g_{i\alpha}$, $\lambda_i$ are constants and $l_i$ is the angular momentum in the $i$th
channel. The elastic potential, $v_{\alpha\beta}$, is approximated by
a constant matrix. 

The omission of rescatterings within the inelastic channels allows us to
disentangle the latter from the elastic channels and from the pole term. We define the potentials 
$$
V_{00}=-\sum_i \raisebox{-2.7mm}{\epsfig{file=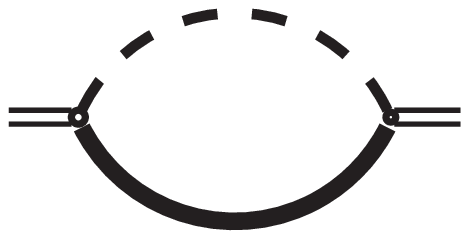,width=16mm}}=-\sum_i \lambda_i^2 J_i,\label{V00def}
$$
\vspace*{-5mm}
$$
V_{\alpha 0}=\raisebox{-3mm}{\epsfig{file=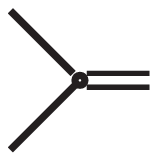,width=9mm}}
-\sum_i\raisebox{-2.7mm}{\epsfig{file=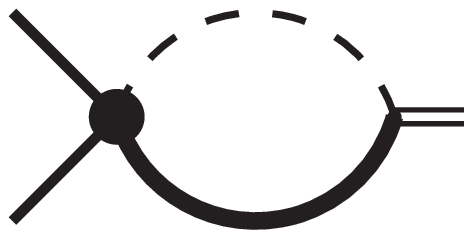,width=16mm}}
=f_\alpha-\sum_i g_{i\alpha}J_i\lambda_i ,\label{Va0}
$$
\vspace*{-5mm}
$$
V_{0\beta}=\raisebox{-3mm}{\epsfig{file=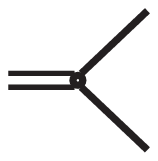,width=9mm}}
-\sum_i\raisebox{-2.7mm}{\epsfig{file=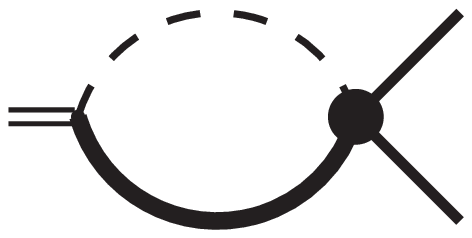,width=16mm}}
=f_\beta-\sum_i \lambda_i J_i g_{i\beta}\label{V0a}
$$
\vspace*{-5mm}
$$
V_{\alpha\beta}=\raisebox{-2.1mm}{\epsfig{file=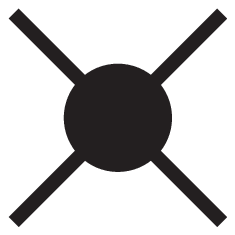,width=7mm}}-\sum_i
\raisebox{-3mm}{\epsfig{file=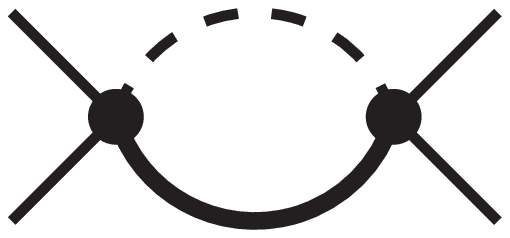,width=17mm}}=v_{\alpha\beta}
-\sum_i g_{i\alpha}J_i g_{i\beta},\label{Vab}
$$
where the thin solid lines, broad solid lines and dashed lines denote heavy-light mesons, heavy mesons, 
and light mesons, respectively, and the double line denotes the pole term. The inelastic loop integral is
\be
J_i=\int |\veq|^{2l_i}\g_i(\veq)\dq\to\frac{i(2\pi)^2}{\sqrt{s}}
m_{th_i^{in}}\mu_i^{in} (k_i^{in})^{2l_i+1},
\ee
where the real part is omitted since it only renormalizes parameters of the interaction; 
$\mu_i^{in}$, $k_i^{in}$, $m_{th_i^{in}}$ are the reduced mass, the relative momentum, and the threshold in the $i$th
inelastic channel, respectively. To disentangle the pole term from the elastic channels we define 
$$
\ds V_{\alpha\beta}^{\rm eff}=v_{\alpha\beta}-G_{\alpha\beta}-V_{\alpha 0}G_0V_{0\beta},~
\ds V_{\alpha 0}^{\rm eff}=V_{\alpha 0}(1+G_0 V_{00}),
$$
where $G_0=1/(M_0-M+V_{00}-i0)$ while the inelastic ``bubble'' operator is
\be
G_{\alpha\beta}\equiv \sum_i
\raisebox{-2.7mm}{\epsfig{file=Vab-2.eps,width=16mm}}
=\sum_ig_{i\alpha}J_ig_{i\beta}.
\label{galbe}
\ee
We arrive, therefore, at a pair of decoupled LSE
\bea
&&t_{\alpha\beta}=V^{\rm eff}_{\alpha\beta}-\sum_\gamma V^{\rm 
eff}_{\alpha\gamma}J_\gamma t_{\gamma\beta},~
t_{\alpha 0}=V^{\rm eff}_{\alpha 0}-\sum_\beta V^{\rm 
eff}_{\alpha\beta}J_\beta t_{\beta 0},\nonumber\\[-4mm]
\label{tabnew}\\[-3mm]
&&J_\alpha=\int \g_\alpha(\vep)d^3p=(2\pi)^2\mu_\alpha(\kappa_\alpha+ik_\alpha)\equiv R_\alpha+iI_\alpha,\nonumber
\eea
with $\mu_\alpha$ and $k_\alpha$ being the reduced mass and the relative momentum in the $\alpha$'s elastic channel,
respectively, $k_\alpha=\sqrt{2\mu_\alpha(M-m_{th_\alpha})+{i \epsilon}}$,
where $m_{th_\alpha}$ is the position of the $\alpha$th elastic threshold.
We reduced the entire problem to Eqs.~(\ref{tabnew}). Thus, independent of the number of 
inelastic channels the solution of these implies only the inversion of
matrices as small as $N_e\times N_e$, where typically $N_e=2$ (cf. the explicit example below). 
The transitions to inelastic channels follow from the solutions to Eqs.~(\ref{tabnew}) straightforwardly, without the 
need to solve another scattering equation.
Therefore, the proposed approach drastically simplifies the combined analysis
of experimental data. In particular, adding a further inelastic channel changes the final expressions only marginally.
Since the  approach is based on a LSE, unitarity is preserved
automatically and all imaginary parts are linked to observable rates.

In order to solve Eqs.~(\ref{tabnew}) we proceed stepwise, analogous to the two-potential
formalism~\cite{twopotform,Hanhart:2012wi}. Our starting point is a convenient parametrization for $\tv$,
the solution of the LSE $\tv=v-vS\tv$, where $v$ is the direct interaction potential in the elastic channels.
The coupling to the inelastic channels is then switched on, and a LSE for the potential
$w=v-G$, where the matrix $G$ was defined in Eq.~(\ref{galbe}),
is solved with the result
\be
\tw=\tv+\psi[{\cal G}-G^{-1}]^{-1}\bar{\psi},
\label{ttt}
\ee
where the dressed vertices and the matrix ${\cal G}$ are
\bea
&\psi_{\alpha\beta}=\delta_{\alpha\beta}\chi_\alpha-\tv_{\alpha\beta}
J_\beta,\quad 
\bar{\psi}_{\alpha\beta}=\delta_{\alpha\beta}\chi_\alpha-
J_\alpha\tv_{\alpha\beta},&\nonumber\\[-2mm]
\label{psi12}\\[-2mm]
&{\cal 
G}_{\alpha\beta}=J_\alpha\psi_{\alpha\beta}=\bar{\psi}_{\alpha\beta}J_\beta=\delta_{\alpha\beta}J_\alpha-J_\alpha\tv_{
\alpha\beta}J_\beta.&\nonumber
\eea

Finally, when the coupling to the pole term is included as well, the
formalism of \cite{Baru:2010ww,Hanhart:2011jz} can be used to yield
$$
t_{\alpha\beta}
=\tw_{\alpha\beta}+\frac{\phi_\alpha\bar{\phi}_\beta}{M-M_0+{\cal G}_0},\quad 
t_{\alpha 0}=\frac{M-M_0}{M-M_0+{\cal G}_0}\phi_\alpha,
\label{ta0sol}
$$
where
\beas
\phi_\alpha=
V_{\alpha 0}-\sum_\beta \tw_{\alpha\beta}J_\beta V_{\beta 0},\quad
\bar{\phi}_\alpha
=V_{0\alpha}-\sum_\beta V_{0\beta}J_\beta\tw_{\beta\alpha},\nonumber\\[-2mm]
{\cal G}_0=\sum_i \lambda_i^2 J_i+\sum_\alpha V_{0\alpha}J_\alpha\phi_\alpha
=\sum_i \lambda_i^2 J_i+\sum_\alpha\bar{\phi}_\alpha J_\alpha V_{\alpha 0}.
\nonumber
\eeas

The $t$ matrix $t_{\alpha i}$ is fully determined by $t_{\alpha\beta}$ and
$t_{\alpha 0}$,
\be
t_{\alpha i}=\left(g_{i\alpha}+\frac{t_{\alpha 0}\lambda_i}{M-M_0}
-\sum_{\beta}t_{\alpha\beta}R_\beta g_{i\beta}\right)(k_i^{in})^{l_i}.
\label{taisol}
\ee

Since our knowledge of most resonance properties comes from production experiments, we build the production
amplitude in the elastic or inelastic channel $x$ as
\be
{\cal M}_x=-\sum_\beta\raisebox{-8.5mm}{\epsfig{file=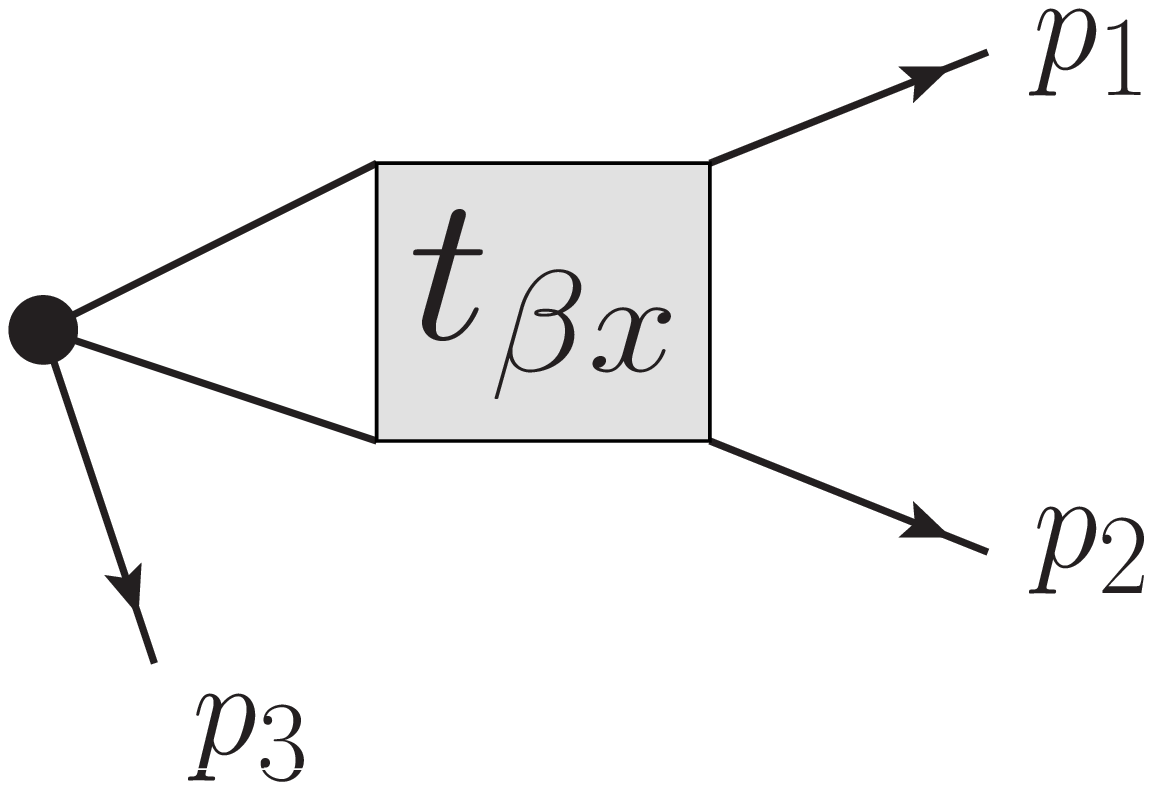,width=0.14\textwidth}}
=-\sum_\beta{\cal F}_\beta J_\beta t_{\beta x},
\label{Mei}
\ee
where it was assumed that the production proceeds through the $N_e$ pointlike elastic sources ${\cal F}_\alpha$. We
also assumed that the elastic $t$ matrix possesses poles near threshold(s) and, therefore, the
Born term in the elastic amplitude was neglected.
The differential production rate can be obtained by integrating
the standard expression for the three-body decay \cite{Agashe:2014kda} in the invariant mass $m_{23}^2$, neglecting the
FSI with
particle 3. Then
\be
\frac{d\Br_x}{dM}=
\frac{|{\cal M}_x|^2p_3k_x}{32\pi^3M_{\rm tot}^2\Gamma_{\rm tot}},\quad M\equiv m_{12}=\sqrt{s},
\label{db1}
\ee
where $k_x$ is the c.m. momentum of particles 1 and 2.
The allowed parameter range for $M$ is given by
$M_{\rm min}=m_1+m_2$ and $M_{\rm max}=M_{\rm
tot}-m_3$.

It is convenient to introduce new parameters $\Lambda ={\cal F}_1^2$ and 
$\xi_{\alpha}={\cal F}_{\alpha}/{\cal F}_1$, 
where the sources ${\cal F}_\alpha$ were redefined to absorb the slow function of energy
$R_\alpha$ and the constant factors from Eq.~(\ref{db1}). Since for all elastic channels the range of forces is
described by
the same physics, it is natural to use $R_\alpha=(2\pi)^2\mu_\alpha\kappa$.
Then the elastic and inelastic differential rates
\bea
&\ds\frac{d\Br_\alpha^e}{dM}=\Lambda \Bigl|\sum_\beta \xi_\beta t_{\beta\alpha}\Bigr|^2p_3k_\alpha,&\label{BreBB0}\\
&{\ds\frac{d\Br_i^{in}}{dM}=\Lambda \Bigl|\sum_\alpha \xi_\alpha
t_{\alpha i}\Bigr|^2p_3k^{in}_i}&\label{Brin0}
\eea
are described by the following set of parameters:
\be
\Lambda ,~\xi_\alpha,~f_\alpha,~\lambda_i,~g_{i\alpha},~M_0,~\kappa,~\tv.
\label{set1}
\ee

\section{Line shapes of the $Z_b(10610)$ and $Z_b(10650)$ states}
\begin{figure*}[t]
\centerline{
\epsfig{file=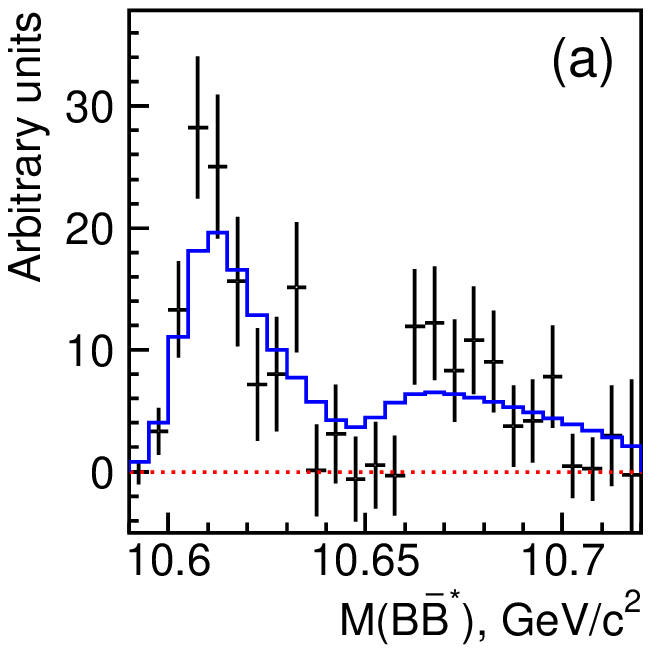,width=0.2\textwidth}
\epsfig{file=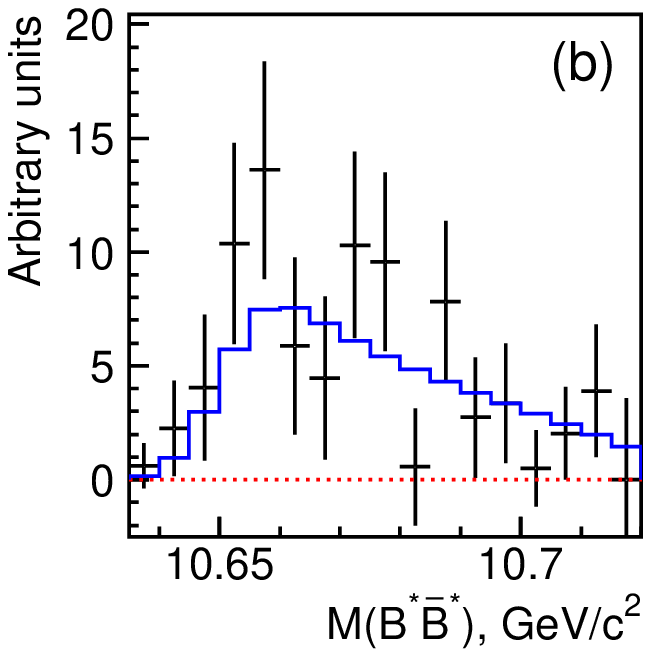,width=0.2\textwidth}
\epsfig{file=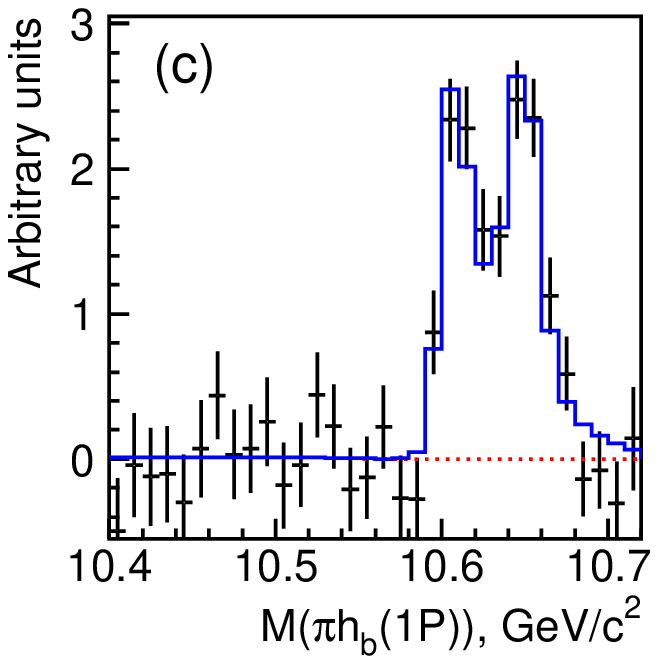,width=0.2\textwidth}
\epsfig{file=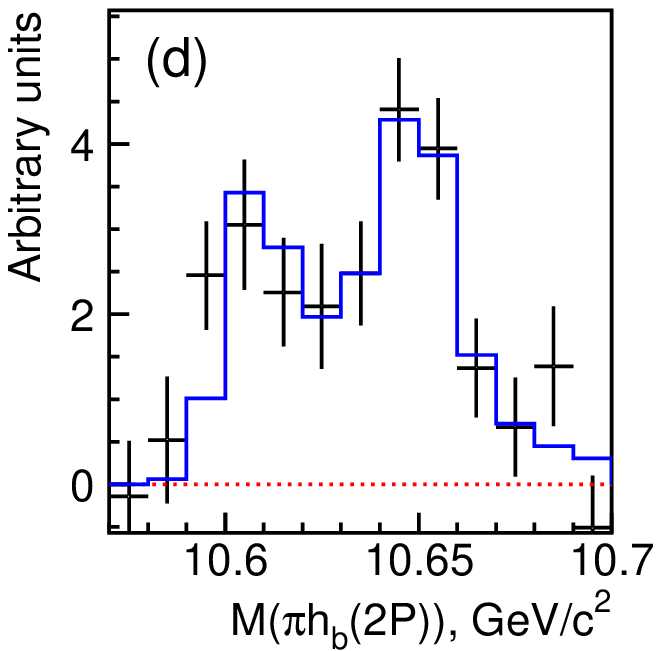,width=0.2\textwidth}
\epsfig{file=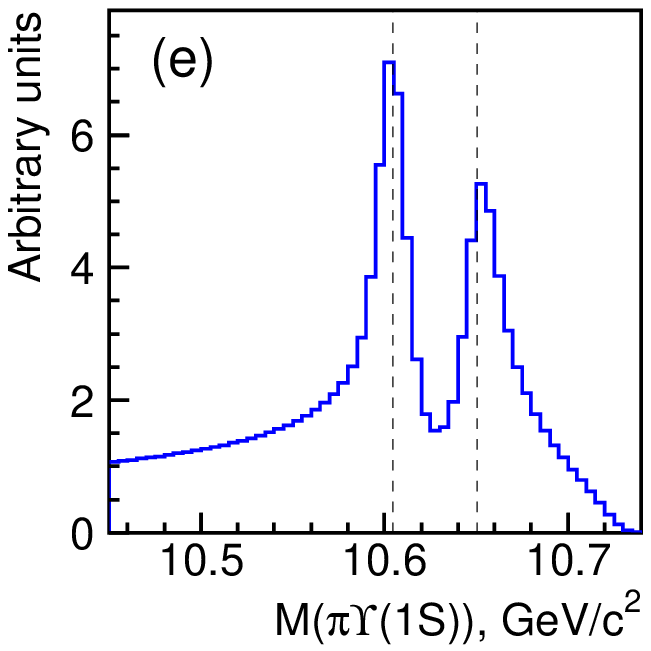,width=0.2\textwidth}
}
\caption{(a)--(d) Fitted line shapes of the $Z_b(10610)$ and $Z_b(10650)$ states in the $B^{(*)}\bar{B}^*$ and $\pi
h_b(mP)$
($m=1,2$) channels and (e) the predicted line shape in the $\pi\Upsilon(1S)$ channel 
[plots for $\pi\Upsilon(2S)$ and $\pi\Upsilon(3S)$ look similar and are not shown). 
Vertical lines indicate $B\bar{B}^*$ and $B^*\bar{B}^*$ thresholds. 
Fit results are given in Eq.~(\ref{params}].}\label{fig:lines}
\end{figure*}

As an application for our approach, we consider $1^{+-}$
$Z_b(10610)$ and $Z_b(10650)$ states residing near the $B\bar{B}^*$ and $B^*\bar{B}^*$ thresholds, respectively,
that are produced in $\Upsilon(5S)$ decays $\Upsilon(5S)\to \pi Z_b$ and that are seen in seven decay channels:
$Z_b\to B\bar{B}^*$, $B^*\bar{B}^*$, $\pi\Upsilon(nS)$ ($n=1,2,3$), and $\pi h_b(mP)$
($m=1,2$) \cite{Belle:2011aa,Adachi:2012cx}. 
The quantum numbers of the final quarkonia fix the angular momenta of the inelastic final states in Eq.~(\ref{Brin0}) to
$l=0$ for all $\pi\Upsilon(nS)$ final states and to $l=1$ for the $\pi h_b(mP)$ final states.

The fact that the $b$-quark mass $m_b\gg \Lambda_{\rm QCD}$ allows us to use heavy-quark spin symmetry (HQSS) to
reduce the number of parameters. If the wave functions of negative-parity heavy-light $B$ mesons
are taken in the form (the charge conjugation is defined as ${\hat C}M=\bar{M}$) $B=0^-_{q \bar b}$, $\bar{B}=0^-_{b
\bar q}$, $B^*=1^-_{q \bar b}$, and $\bar{B}^*=-1^-_{b \bar q}$, then the Fierz transformation yields the following
$1^{+-}$ combinations:
\beas
&&|B\bar{B}^*\rangle_{1^{+-}}=-\frac{1}{\sqrt 2}\Bigl[(1^-_{b \bar b}\otimes 0^-_{q \bar q})_{S=1}+
(0^-_{b \bar b}\otimes 1^-_{q \bar q})_{S=1}\Bigr]\nonumber,\\
&&|B^*\bar{B}^*\rangle_{1^{+-}}=\frac{1}{\sqrt{2}}\Bigl[(1^-_{b \bar b}\otimes 0^-_{q \bar
q})_{S=1}-(0^-_{b \bar b}\otimes 1^-_{q \bar q})_{S=1}\Bigr], \nonumber
\eeas
which imply that \cite{Bondar:2011ev,Voloshin:2011qa}
\be
\frac{g_{[\pi\Upsilon(nS)][B^*\bar{B}^*]}}{g_{[\pi\Upsilon(nS)][B\bar{B}^*]}}=-1,\quad 
\frac{g_{[\pi h_b(mP)][B^*\bar{B}^*]}}{g_{[\pi h_b(mP)][B\bar{B}^*]}}=1,
\label{HSconstr}
\ee
where the total angular momentum of the light-quark contribution in the latter
case is to be provided by one unit of angular momentum that is explicitly accounted
for in Eq.~(\ref{Brin0}). Once the elastic channels $B\bar{B}^*$ and $B^*\bar{B}^*$ are
produced in the decays of the $\Upsilon(5S)$, the ratio of the sources $\xi$ is subject to the same heavy-quark
constraint,
\be
\xi=\frac{g_{[\pi\Upsilon(5S)][B^*\bar{B}^*]}}{g_{[\pi\Upsilon(5S)][B\bar{B}^*]}}=-1.
\label{xi}
\ee
In the same limit, the direct interaction in the $B^{(*)}\bar{B}^*$ system
can be parametrized in terms of only two parameters, $\gamma_s$ and $\gamma_t$, which are related to
the contact potentials used in \cite{Nieves:2012tt} as
$\gamma_s^{-1}=(2\pi)^2\mu(C_{0a}+C_{0b})$ and $\gamma_t^{-1}=(2\pi)^2\mu(C_{0a}-3C_{0b})$ 
($\mu_1\approx\mu_2\equiv\mu$). Then \cite{Hanhart:2011jz}
$$
t^v=\frac{1}{(2\pi)^2\mu}\frac{1}{\mbox{Det}}
\left(
\begin{array}{cc}
\frac12(\gamma_s+\gamma_t)+ik_2&\frac12(\gamma_t-\gamma_s)\\
\frac12(\gamma_t-\gamma_s)&\frac12(\gamma_s+\gamma_t)+ik_1
\end{array}
\right),
\label{tv}
$$
with $\mbox{Det}=\gamma_s\gamma_t-k_1k_2+(i/2)(\gamma_s+\gamma_t)(k_1+k_2)$.

The bare pole is included in the formalism in order to provide more flexibility in
the fitting process---in particular, it allows one to have two poles
near the threshold even in the single-channel case. However, it should be omitted if its presence is not requested by
the
data. Thus, since we get a very good fit even without the bare pole, we refrain from its inclusion, thus setting
$f_\alpha=\lambda_i=0$ and $M_0\to\infty$ in all formulas. As an experimental input we use
\begin{itemize}
\item background-subtracted and efficiency-corrected distributions in $M$ for the $B^{(*)}\bar{B}^*$ and $\pi h_b(nP)$
channels \cite{Adachi:2012cx,Belle:2011aa} with floating normalization in each channel;
\item ratios of total branching fractions, $\Br^e_{B\bar{B}^*}/\Br^e_{B^*\bar{B}^*}$, 
$\Br^{in}_i/\Br^e_{B^*\bar{B}^*}$ \cite{Adachi:2012cx,Belle:2011aa,Adachi:2011ji,Garmash:2014dhx},
where index $i$ runs over all five inelastic channels $\pi\Upsilon(nS)$ and $\pi h_b(mP)$.
\end{itemize}
We do not use the information on the $Z_b^{(\prime)}$ line shapes in the $\pi\Upsilon(nS)$ channels, since in the
one-dimensional fit it is not possible to correctly take into account the interference with the nonresonant continuum,
which is significant in the $\Upsilon(5S)\to\pi^+\pi^-\Upsilon(nS)$ transitions. Inclusion of the $\pi\Upsilon(nS)$ line
shapes would require a multidimensional analysis, which is beyond the scope of this Letter; however, it is
straightforward from the theoretical point of view. In order to come to a converging fit we are, therefore, forced to
impose that $g_{[\pi\Upsilon(nS)][B\bar B^*]}=-g_{[\pi\Upsilon(nS)] [B^*\bar B^*]}$ for $n=1,2,3$, as given by
Eq.~(\ref{HSconstr}). Meanwhile, we leave $g_{[\pi\Upsilon(5S)][B\bar B^*]}$ and $g_{[\pi\Upsilon(5S)] [B^*\bar B^*]}$
as well as $g_{[\pi h_b(mP)][B\bar B^*]}$ and $g_{[\pi h_b(mP)] [B^*\bar B^*]}$ unconstrained. 
The line shapes in the $\pi\Upsilon(nS)$ channels come out as our prediction.
To take into account the experimental resolution, we convolve all the distributions with a Gaussian function
with $\sigma=6$~MeV. Results of the simultaneous fit are shown in Figs.~\ref{fig:lines}(a)--\ref{fig:lines}(d). The
developed
parametrization provides a very good description of the experimental data, with a confidence level of 76\%. Predicted
line shapes in the $\pi\Upsilon(nS)$ channels look reasonably similar to the experimental data~\cite{Garmash:2014dhx};
an example of such a distribution is shown in Fig.~\ref{fig:lines}(e). It turns out that parameter $\kappa$, defined in
Eq.~(\ref{tabnew}), is practically unconstrained by the fit; thus, we fix it to 1~GeV. From the fit we find
\bea
&\xi=\ds\frac{g_{[\pi\Upsilon(5S)][B^*\bar{B}^*]}}{g_{[\pi\Upsilon(5S)][B\bar{B}^*]}}=-0.84\pm 0.05,&\nonumber\\[1mm]
&\ds\frac{g_{[\pi h_b(1P)][B^*\bar{B}^*]}}{g_{[\pi h_b(1P)][B\bar{B}^*]}}=2.4\pm 0.6,&\label{params}\\[1mm]
&\ds\frac{g_{[\pi h_b(2P)][B^*\bar{B}^*]}}{g_{[\pi h_b(2P)][B\bar{B}^*]}}=2.4\pm 0.7.&\nonumber
\eea
Deviations of these parameters from the predictions of HQSS---see Eqs.~(\ref{HSconstr}) and (\ref{xi})---might be
explained by the close proximity of the
$t$ matrix poles to the thresholds, which can result in an enhancement of
the small explicit symmetry violation caused by $\Lambda_{\rm QCD}/m_b\neq 0$~\cite{Cleven:2011gp}. Another source of
the deviation of $\xi$ from the prediction of the HQSS may stem from a $D$-wave $b\bar{b}$
component as well as from possible non-$b\bar{b}$ components of the
$\Upsilon(5S)$ wave function (cf. the discussion in \cite{Guo:2014qra}).
The importance of the HQSS-breaking contributions for the proper description of 
the $Z_b^{(\prime)}$ line shapes was also stressed in \cite{Mehen:2013mva}. 
It should be noticed that preliminary Belle data on the $B\bar{B}^*$ and $B^*\bar{B}^*$ channels
are used in the current analysis; fit results could change for the final experimental data. 
The inclusion of the information on the $\pi\Upsilon(nS)$ line shapes in a future multidimensional analysis 
will help to improve the accuracy of the determination of the model parameters
and will allow for drawing more firm conclusions about the underlying physics of the spectacular
near-threshold phenomena called $Z_b(10610)$ and $Z_b(10650)$.

\section{Conclusions}\label{sum}

In this Letter we proposed a practical parametrization for the line shapes of the near-threshold state(s). 
Since the approach is based on the LSE for the coupled-channel problem, unitarity and analyticity constraints for
the $t$ matrix are fulfilled automatically. This guarantees that all imaginary parts are included in a self--consistent
way. Since there are good reasons to neglect direct interactions within the inelastic
channels, at least for the systems discussed here, the inelastic channels enter the expressions additively which; this
makes
it particularly easy to extend the inelastic basis. While additional effects such as finite widths of the constituents
and the FSI with the spectator may also play a role and
should be included on top of the interactions considered in this Letter; nevertheless, the gross features of the
coupled-channel problem are captured by the presented model and the parametrization based on it is
expected to be realistic. Finally, we demonstrate the power of the suggested parametrization by the fit to the line
shapes for the bottomoniumlike states $Z_b(10610)$ and $Z_b(10650)$, for which we obtain a  very good description. 
\smallskip

\begin{acknowledgments}
We would like to thank Alexander Bondar, Martin Cleven, Feng-Kun Guo, and Ulf-G. Mei\ss ner 
for valuable discussions. This work is
supported in part by the DFG and the NSFC through funds
provided to the Sino-German CRC 110 ``Symmetries and
the Emergence of Structure in QCD.'' 
R.M. and A.N. are supported by the Russian Science Foundation (Grant No. 15-12-30014).
\end{acknowledgments}

\end{document}